# Improving the Fidelity of deletion

*Satyabrata Adhikari, **B.S.Choudhury

Department of Mathematics,

Bengal Engineering and Science University

Howrah-711103, West Bengal, India.

E-Mail: *satyyabrata@yahoo.com, **binayak12@yahoo.co.in

**Abstract:** In this work we study the quantum deletion machine with two transformers and show that the deletion machine with single transformer performs better than the deletion machine with more than two transformers. We also observe that the fidelity of deletion depends on the blank state used in the deleter and so for different blank state the fidelity is different. Further, we study the Pati-Braunsein deleter with transformer.

## I. Introduction

Quantum mechanics prevents cloning (copying) of an arbitrary quantum state, nevertheless approximate cloning is possible. Description of various types of cloning machines may be obtained in [1,2,3,4]. An interesting application of quantum cloning is the broadcasting of entanglement proposed by Buzek et.al. [11]. The complementary theory of "quantum no-cloning theorem" is the "quantum no-deleting" principle [5]. It states that linearity of quantum theory forbids deleting unknown quantum state against a copy. Quantum deletion [6] is more like reversible 'uncopying' of an unknown quantum state. The corresponding no-deleting principle does not prohibit us from constructing the approximate deleting machine [5,6]. When memory in a quantum computer is scare, quantum deleting may play an important role. J. Feng et.al. [7] showed that each of two copies of non-orthogonal and linearly independent quantum states can be probabilistically deleted by a general unitary-reduction operation. Like universal quantum cloning machine, D'Qiu [8] also constructed a universal deletion machine but unfortunately the machine was found to be non-optimal in the sense of fidelity. Recently in [10], we studied the deletion of one imperfect copy obtained from approximate quantum cloning machines. In fact construction of universal optimal deletion machine remains an open problem. Recently, we designed a universal quantum deletion machine in an unconventional way that improves the fidelity of deletion from 0.5 and takes it to 0.75 in the limiting sense [12]. The universal deletion machine consists of two unitary operators namely, deleter and transformer. Deleter takes two identically prepared input qubits and then deletes a state of the qubit in a usual manner. After completion of the function of deleter, the resulting state enters into the transformation chamber. The transformation chamber is nothing but a unitary operator T [9] which transforms the resulting state. Then after passing through the whole procedure, we observed that the deleting machine deletes a qubit in the second mode with fidelity 0.75 and retain the qubit in the first mode with average fidelity 0.77. Therefore, we can notice that the fidelity of deletion improves to some extent. Thus our main aim of this work is to investigate the reason behind the increment of the fidelity of deletion. The fidelity of deletion can increase because of the following two reasons given below:

(i) Introduction of the second unitary operator T for transformation purpose.

(ii) Application of suitable deletor for deletion purpose.

Our objective here is to study the above-mentioned reasons for enhancement of the fidelity of deletion. In section II, we revisit the quantum deletion machine with single transformer. In section III, we use two transformers in the quantum deleting machine and study whether there is any enhancement in the fidelity of deletion. In section IV, we use transformer in the Pati-Braunstein (P-B) deleter and hence observe the good performance of the P-B quantum deleting machine. We show that the deletion machine consists of P-B deleter and transformer reduces to universal deletion machine and not only that, the fidelity of deletion increases to 0.85.

**II. Revisit of quantum deletion machine with one transformer**

In this section, we revisit the quantum deletion machine with one transformer. This type of deletion machine requires two vital parts: deleter and transformer.

**1.Deleter:** Deleter is nothing but a unitary transformation (deletion transformation), which is used to delete a copy from among two given copy of an unknown quantum state.

A unitary transformation U which describes a deleter is given below:

$$U|00\rangle_{ab}|A\rangle_c \to |0\rangle_a|\Sigma\rangle_b|A_0\rangle_c + \left[|0\rangle_a|1\rangle_b + |1\rangle_a|0\rangle_b\right]|B_0\rangle_c \qquad (2.1)$$

$$U|01\rangle_{ab}|A\rangle_c \to |0\rangle_a|\Sigma_\perp\rangle_b|D_0\rangle_c + |1\rangle_a|0\rangle_b|C_0\rangle_c \qquad (2.2)$$

$$U|10\rangle_{ab}|A\rangle_c \to |1\rangle_a|\Sigma\rangle_b|D_0\rangle_c + |0\rangle_a|1\rangle_b|C_0\rangle_c \qquad (2.3)$$

$$U|11\rangle_{ab}|A\rangle_c \to |1\rangle_a|\Sigma_\perp\rangle_b|A_1\rangle_c + \left[|0\rangle_a|1\rangle_b + |1\rangle_a|0\rangle_b\right]|B_1\rangle_c \qquad (2.4)$$

where $|A\rangle$ is the initial and $|A_i\rangle, |B_i\rangle, |C_j\rangle, |D_j\rangle (i=0,1; j=0)$ are the final machine state vector. $|\Sigma\rangle$ is some standard state and $|\Sigma_\perp\rangle$ denote a state orthogonal to $|\Sigma\rangle$.

**2. Transformer:** It is described by a unitary transformation T. It is used in the deletion machine in such a way so as to increase the fidelity of deletion and minimize the distortion of the undeleted qubit. The unitary operator T [9,12] is defined by $T = |\psi^+\rangle\langle 00| + |11\rangle\langle 01| + |\psi^-\rangle\langle 10| + |00\rangle\langle 11|$

Where $|\psi^\pm\rangle = (1/\sqrt{2})(|01\rangle \pm |10\rangle)$.

Therefore the 'deletion machine with one transformer' [12] deletes a qubit with fidelity of deletion

$$F_1 = \frac{1}{2}\left[1 + m_1 m_2 - \frac{m_1^2 - m_2^2}{\sqrt{2}}\right] \text{ as } \lambda \to \frac{1}{2}, \text{ (for real } m_1 \text{ and } m_2\text{)} \qquad (2.5)$$

Where '$\lambda$' is the machine parameter and the standard blank state is given by

$|\Sigma\rangle = m_1|0\rangle + m_2|1\rangle$ and $|\Sigma_\perp\rangle = -m_2^*|0\rangle + m_1|1\rangle$ with $m_1$ real and $m_2$ complex such that

$m_1^2 + |m_2|^2 = 1$. Therefore, $\langle 0|\Sigma\rangle = \langle \Sigma|0\rangle = m_1$, $\langle 1|\Sigma\rangle = m_2$, $\langle \Sigma|1\rangle = m_2^*$.

Equation (2.5) shows that the fidelity of deletion remains the same for all input state and it depends only on the blank state. Since the limiting fidelity of deletion of the deletion machine with one transformer depends on the parameters $m_1$ and $m_2$ so the variation of the limiting fidelity of deletion with $m_1$ and $m_2$ is studied and given in the table below:

TABLE-I

| $m_1^2$ | $m_2^2$ | $m_1^2 - m_2^2$ | Limiting Fidelity $(F_1) = \frac{1}{2}\left[1 \pm \frac{\sqrt{1-(m_1^2-m_2^2)^2}}{2} - \frac{m_1^2-m_2^2}{\sqrt{2}}\right]$ (according as $m_1 m_2 > 0$ or $m_1 m_2 < 0$)  (upto two significant figures) |
|---|---|---|---|
| 0.0 | 1.0 | -1.0 | 0.85 |
| 0.1 | 0.9 | -0.8 | 0.93 or 0.63 |
| 0.2 | 0.8 | -0.6 | 0.91 or 0.51 |
| 0.3 | 0.7 | -0.4 | 0.87 or 0.41 |
| 0.4 | 0.6 | -0.2 | 0.81 or 0.32 |
| 0.5 | 0.5 | 0.0 | 0.75       (obtained in [12]) |
| 0.6 | 0.4 | 0.2 | 0.67 or 0.18 |
| 0.7 | 0.3 | 0.4 | 0.58 or 0.12 |
| 0.8 | 0.2 | 0.6 | 0.48 or 0.08 |
| 0.9 | 0.1 | 0.8 | 0.36 or 0.06 |
| 1.0 | 0.0 | 1.0 | 0.14 |

Therefore, we can observe from the above table that if the product of the parameter of the blank state is negative (i.e. when either $m_1$ or $m_2$ is negative) then the deletion machine deletes the state with lower fidelity of deletion but if the product of the parameter of the blank state is positive (i.e. when both $m_1$ and $m_2$ is negative or positive) then the deletion machine performs well in the sense of high fidelity of deletion. In this work we discuss the quantum deletion machine with one transformer for various values of the parameters $m_1$ and $m_2$ and find that the quantum deletion machine really works well for some blank state and with the help of those blank states, quantum deletion machine deletes a quantum state with fidelity of deletion higher than 0.75.

### III. Quantum deletion machine with two transformers

In this section, we study the quantum deletion machine (2.1-2.4) with the addition of two transformers. Since the introduction of the transformer increases the fidelity of deletion so one may expect that the application of transformer more than one time increases the fidelity of deletion further and therefore there may exist a threshold number of the transformers (i.e. maximum number of transformers) whose application on the deleted state increases the fidelity of deletion to its optimal value. But we will show that this is not necessarily true.

Let $|\psi\rangle = \alpha|0\rangle + \beta|1\rangle$ with $\alpha^2 + |\beta|^2 = 1$ be any unknown quantum state.

Where $\alpha$ is real and $\beta$ is complex.

The modified deletion machine with two transformers deletes one of the copies of an input state $|\psi\rangle|\psi\rangle$ and then after transforming the deleted qubit, the final output state of the deletion machine described by the density operator $\rho_{12c}^{out} = (T)^2 |\chi_d^{out}\rangle_{12c} \langle\chi_d^{out}| (T^t)^2$ where $|\chi_d^{out}\rangle$ represent a state after passing through the deleter (2.1-2.4). Since we are interested to see the performance of the deletion machine with two transformers in the sense that how well it deletes a qubit, we only consider the state of the qubit in mode 2. Therefore, the reduced density operator describe the state of the qubit in

mode '2' is given by $\rho_2^{out}$. But as $\lambda \to \frac{1}{2}$, $\rho_2^{out} \to \rho_2'^{out}$,

$$\rho_2'^{out} = \frac{5}{8}|0\rangle\langle 0| + \frac{1}{4}\left(\frac{1}{\sqrt{2}}-1\right)|0\rangle\langle 1| + \frac{1}{4}\left(\frac{1}{\sqrt{2}}-1\right)|1\rangle\langle 0| + \frac{3}{8}|1\rangle\langle 1| \qquad (3.1)$$

Equation (3.1) shows that the state described by the density operator $\rho_2'^{out}$ is input state independent. Therefore, whatever be the input quantum state, after passing through the deleter and two transformers, the resulting output state remains same for any arbitrary input state.

The limiting fidelity of deletion is given by

$$F = \langle \Sigma' | \rho_2'^{out} | \Sigma' \rangle, \text{ where } |\Sigma'\rangle = (1/\sqrt{2})(|\Sigma\rangle + |\Sigma_\perp\rangle) \qquad (3.2)$$

Hence, $F = \frac{1}{2}\left[\frac{5}{8}(1 - m_1 m_2^* - m_1 m_2) + \frac{1}{4}\left(\frac{1}{\sqrt{2}}-1\right)(2m_1^2 - (m_2^*)^2 - m_2^2) + \frac{3}{8}(1 + m_1 m_2^* + m_1 m_2)\right]$ (3.3)

Equation (3.3) shows that the fidelity of deletion varies as $m_1$ and $m_2$ i.e. the limiting fidelity depends on the blank state used in the deleter but not on the arbitrary input state. Also equation (3.3) gives the fidelity of deletion for any arbitrary blank state.

In particular, if we assume $m_1$ and $m_2$ to be real, then the variation of fidelity with the amplitudes of the blank state $m_1$ and $m_2$ is given below:

TABLE-II

| $m_1^2$ | $m_2^2$ | $m_1^2 - m_2^2$ | Limiting Fidelity (F) = $\frac{1}{2}\left[1 \mp \frac{\sqrt{1-(m_1^2 - m_2^2)^2}}{4} + \left(\frac{1}{2}\right)\left(\frac{1}{\sqrt{2}}-1\right)(m_1^2 - m_2^2)\right]$ |
|---|---|---|---|
| | | | (according as $m_1 m_2 > 0$ or $m_1 m_2 < 0$)   (upto two significant figures) |
| 0.0 | 1.0 | -1.0 | 0.57 |
| 0.1 | 0.9 | -0.8 | 0.48 or 0.63 |
| 0.2 | 0.8 | -0.6 | 0.44 or 0.64 |
| 0.3 | 0.7 | -0.4 | 0.41 or 0.64 |
| 0.4 | 0.6 | -0.2 | 0.39 or 0.63 |
| 0.5 | 0.5 | 0.0 | 0.37 |
| 0.6 | 0.4 | 0.2 | 0.36 or 0.60 |
| 0.7 | 0.3 | 0.4 | 0.35 or 0.58 |
| 0.8 | 0.2 | 0.6 | 0.35 or 0.55 |

| 0.9 | 0.1 | 0.8 | 0.36 or 0.51 |
| 1.0 | 0.0 | 1.0 | 0.42 |

On the contrary of the quantum deletion machine with a single transformer, we observe here that if the product of the parameter of the blank state is negative (i.e. when either $m_1$ or $m_2$ is negative) then the deletion machine deletes the state with fidelity of deletion higher than the case when the product of the parameter of the blank state is positive (i.e. when both $m_1$ and $m_2$ is negative or positive). Also we note that the deletion machine with two transformers deletes a state with fidelity of deletion 0.37 when we use the blank state with parameter $m_1 = m_2 = 0.5$. In addition to this, If we compare the quantum deletion machine with two transformer with the quantum deletion machine with a single transformer then we find that the deletion machine with single transformer works better when the product of the amplitudes of the blank state $m_1$ and $m_2$ is positive while the deletion machine with two transformer works better when the product of the parameters $m_1$ and $m_2$ is negative.

## IV. P-B deleting machine with transformer

In this section, we study the conditional P-B deleting machine with addition of unitary operator called transformer T. We will show in this section that the addition of transformer in the quantum deletion machine increases the fidelity of deletion and makes the fidelity of deletion input state independent.

The conditional P-B deleting transformation is given by

$$|0\rangle|0\rangle|A\rangle \rightarrow |0\rangle|\Sigma\rangle|A_0\rangle \qquad (4.1)$$

$$|1\rangle|1\rangle|A\rangle \rightarrow |1\rangle|\Sigma\rangle|A_1\rangle \qquad (4.2)$$

$$|0\rangle|1\rangle|A\rangle \rightarrow |0\rangle|1\rangle|A\rangle \qquad (4.3)$$

$$|1\rangle|0\rangle|A\rangle \rightarrow |1\rangle|0\rangle|A\rangle \qquad (4.4)$$

Where $|A\rangle, |A_0\rangle, |A_1\rangle$ are orthonormal machine state vectors.

The deletion machine (P-B deleter + Transformer) deletes one copy from two copies of the input state $|\psi\rangle = \alpha|0\rangle + \beta|1\rangle$ with $\alpha^2 + |\beta|^2 = 1$ and the resulting output state from the deletion machine is

given by $\rho_{12c}^{out} = T |\psi_d^{out}\rangle_{12c} \langle\psi_d^{out}| T^t$, where $|\psi_d^{out}\rangle_{12c}$ denotes the state after passing through the P-B deleter.

The reduced density operator in mode '2' is given by

$$\rho_2 = Tr_{1c}\left(T |\psi_d^{out}\rangle_{12c} \langle\psi_d^{out}| T^t\right)$$

$$= |0\rangle\langle 0|\left(\frac{\alpha^4 m_1^2}{2} + \frac{\alpha^2 |\beta|^2}{2} + \frac{|\beta|^4 m_1^2}{2} + |\beta|^4 |m_2|^2\right) + |0\rangle\langle 1|\left(\frac{\alpha^4 m_1 m_2^*}{\sqrt{2}} - \frac{\alpha^2 |\beta|^2}{\sqrt{2}} + \frac{|\beta|^4 m_1 m_2}{\sqrt{2}}\right)$$

$$+ |1\rangle\langle 0|\left(\frac{\alpha^4 m_1 m_2}{\sqrt{2}} - \frac{\alpha^2 |\beta|^2}{\sqrt{2}} + \frac{|\beta|^4 m_1 m_2^*}{\sqrt{2}}\right) + |1\rangle\langle 1|\left(\frac{\alpha^4 m_1^2}{2} + \frac{3\alpha^2 |\beta|^2}{2} + \frac{|\beta|^4 m_1^2}{2} + \alpha^4 |m_2|^2\right) \quad (4.5)$$

Now to see how well our deleting system deletes a qubit, we have to calculate the fidelity of deletion defined by $F_2 = \langle\Sigma|\rho_2|\Sigma\rangle$

If we assume $m_1$ and $m_2$ to be real, then

$$F_2 = m_1^2\left[\frac{m_1^2}{2} + \alpha^2|\beta|^2\left(\frac{1-2m_1^2}{2}\right) + |\beta|^4 m_2^2\right] + m_1 m_2\left[\frac{m_1 m_2}{\sqrt{2}} - \alpha^2|\beta|^2\left(\frac{2m_1 m_2 + 1}{\sqrt{2}}\right)\right]$$

$$+ m_1 m_2\left[\frac{m_1 m_2}{\sqrt{2}} - \alpha^2|\beta|^2\left(\frac{2m_1 m_2 + 1}{\sqrt{2}}\right)\right] + m_2^2\left[\frac{m_1^2}{2} + \alpha^2|\beta|^2\left(\frac{3-2m_1^2}{2}\right) + \alpha^4 m_2^2\right] \quad (4.6)$$

If $m_1 = \frac{1}{\sqrt{2}}$ and $m_2 = \frac{-1}{\sqrt{2}}$, then $F_2 = \frac{1}{2} + \frac{1}{2\sqrt{2}} = 0.85$ (*approx.*) \quad (4.7)

Equation (4.7) shows that there exist a blank state for which the fidelity of deletion is input state independent and also its value approaches the optimal cloning fidelity, which we expect from our universal deletion machine. Therefore, the advantage of using the transformer in the quantum deletion machine with P-B deleter is that the machine deletes a qubit with fidelity 0.85 which remains same for all input state. In addition to this, we can observe that the average fidelity of deletion (0.85) for a deletion machine with P-B deleter and transformer is greater than the average fidelity (0.83) for a deletion machine consists of only P-B deleter.

**Conclusion**

In this work we introduce the "quantum deletion machine with two transformers" and compare it with the deletion machine with single transformer. Here we observe an important fact that the quantum deletion machine with single transformer need not be always performs better than the quantum deletion machine with more than one transformer. In some cases quantum deletion machine with two transformers deletes a state of the qubit with better fidelity of deletion. Also we note that the fidelity of deletion depends on the parameter of the blank state, hence the blank state also take part as a key component of the deletion machine.

The quantum deletion machine with single transformer deletes a qubit with highest fidelity $F_1 = 0.93$, when $m_1 = 0.31$ and $m_2 = 0.94$

$= 0.63$, when $m_1 = 0.31$ and $m_2 = -0.94$ or $m_1 = -0.31$ and $m_2 = 0.94$.

The quantum deletion machine with two transformers deletes a qubit with highest fidelity

$F = 0.57$, when $m_1 = 0.0$ and $m_2 = 1.0$

$= 0.63$, when $m_1 = 0.31$ and $m_2 = -0.94$ or $m_1 = -0.31$ and $m_2 = 0.94$

$= 0.63$, when $m_1 = 0.63$ and $m_2 = -0.77$ or $m_1 = -0.63$ and $m_2 = 0.77$

We also study the quantum deletion machine consisting of deleter obeying Pati-Braunstein transformation rule and transformer. It is a well-known fact that the P-B conditional deletion machine is state dependent and the average fidelity of deletion is 0.83 but here we find that the fidelity of deletion of the P-B conditional deletion machine with single transformer is input state independent for a given blank state and the value of the fidelity of deletion is 0.85 (approx.).

**Acknowledgement:**

The present work is supported by CSIR project no. F.No.8/3(38)/2003-EMR-1, New Delhi. The support is gratefully acknowledged.